\documentclass[conference]{IEEEtran}
\IEEEoverridecommandlockouts

\usepackage{cite}
\usepackage{amsmath,amssymb,amsfonts}
\usepackage{algorithmic}
\usepackage{graphicx}
\usepackage{textcomp}
\usepackage{xcolor}
\usepackage{tikz}
\usepackage{booktabs}
\usepackage[font=small]{caption}
\usepackage{subcaption}
\usepackage{hyperref}

\newcommand\copyrighttext{%
  \footnotesize \textcopyright 2021 IEEE. Personal use of this material is permitted.
  Permission from IEEE must be obtained for all other uses, in any current or future
  media, including reprinting/republishing this material for advertising or promotional
  purposes, creating new collective works, for resale or redistribution to servers or
  lists, or reuse of any copyrighted component of this work in other works.}
\newcommand\copyrightnotice{%
\begin{tikzpicture}[remember picture,overlay]
\node[anchor=south,yshift=10pt] at (current page.south) {\fbox{\parbox{\dimexpr\textwidth-\fboxsep-\fboxrule\relax}{\copyrighttext}}};
\end{tikzpicture}%
}

\def\BibTeX{{\rm B\kern-.05em{\sc i\kern-.025em b}\kern-.08em
    T\kern-.1667em\lower.7ex\hbox{E}\kern-.125emX}}

\IEEEaftertitletext{\vspace{-1.6\baselineskip}}

\makeatletter

\graphicspath{{./fig/}}
\begin{document}

\title{Steady State Modeling for Variable Frequency AC Power Flow
\thanks{This material is based upon work supported by the New York Power Authority (NYPA), the New York State Energy Research and Development Authority (NYSERDA), and the National Science Foundation Graduate Research Fellowship Program under Grant No. DGE-1747503.}
}

\author{\IEEEauthorblockN{David Sehloff and Line Roald}
\IEEEauthorblockA{%
University of Wisconsin--Madison, Madison, Wisconsin, USA \\
Email: dsehloff@wisc.edu; roald@wisc.edu}
}

\maketitle
\copyrightnotice

\begin{abstract}
Advantages of operating portions of a power system at frequencies different from the standard 50 or 60 Hz have been demonstrated in the low frequency AC (LFAC) and high voltage DC (HVDC) literature. Branches constrained by stability or thermal limits can benefit from increased capacity and flexibility. Since advances in power electronics enable the choice of an operating frequency, tools are needed to make this choice. In order to quantify the advantages as functions of frequency, this paper provides models for steady state calculations with frequency as a variable and validates the modeling assumptions. It then introduces an analytical quantification of the power flow capacity of a transmission branch as a function of frequency, demonstrating different active constraints across the range of frequency. The modeling and power flow calculations are demonstrated for a practical transmission line using manufacturer data. The models presented here provide a foundation for system level studies with variable frequency using optimal power flow.
\end{abstract}


\section{Introduction}
Ever since AC power systems won the initial war of the currents, the use of a single standard frequency has been a central tenet in the planning and control of large power systems. Accordingly, the modeling and analysis techniques employed in these fields have developed based on the assumption of a single, fixed frequency of 50 or 60 Hz. With recent developments in power electronics, this assumption is being undermined by the increasing prevalence of transmission at various frequencies, most notably high voltage DC (HVDC) and the emerging applications of low frequency AC (LFAC).

This creates two needs. First, to analyze system operations at different frequencies, it is necessary to develop a steady state system model which explicitly accounts for frequency dependence in the component parameters. While this model may build upon existing power system models, it is important that the underlying modeling assumptions are validated across the full range of frequencies under consideration. Second, given the above model, it is possible to develop a quantitative understanding of the advantages that can be gained by varying the frequency. In this paper, we aim to fill these needs by discussing the frequency-dependent model parameters, provide a validation of the power flow model and apply it to show how the benefits vary over a range of frequencies.

Many of the advantages of transmission at frequencies lower than 50 Hz were demonstrated in \cite{Wang1996,Funaki2000,meliopoulos2012} with a focus on the advantages of low frequency AC (LFAC) for power flow capacity. The lower branch reactance achieved by lowering the frequency improves the operation of both overhead lines \cite{Wang1996} and cables \cite{Funaki2000} by easing limits related to transient stability, voltage stability, and thermal considerations. For more than a century, low frequencies, especially 16 2/3 and 25 Hz, have been used for transmission of railway traction power \cite{steimel2012,laury2018} and for industrial purposes \cite{Blalock2008}, applications which have benefited from the use of modern power electronics \cite{ABB2018,ABB2012}. Recently, offshore wind \cite{Fischer2012,Olsen2014,Ruddy2016}, and long distance overhead transmission \cite{Wang1996,Wang2006} have emerged as promising applications.

The steady state analysis of networks with nonstandard frequencies has been the subject of recent work. In \cite{Nguyen2016}, Nguyen and Santoso formulated a power flow problem for a 60 Hz network with a single branch at 10 Hz, with frequency conversion at each end. The authors used the $\Pi$ equivalent model for the low frequency branch. In \cite{Nguyen2019}, this power flow model was extended to accommodate networks composed of multiple areas with frequency conversion at the area interfaces, allowing for distinct frequencies, including DC, in each area. In \cite{Nguyen2019opf}, an optimal power flow problem was formulated with multiple areas of distinct frequencies. The frequency was not considered variable in the formulation but fixed before the calculation. This work helped demonstrate the benefits of LFAC and the potential of tools for planning and operating networks with multiple frequencies. However, existing work does not answer what would be the optimal choice of frequency, nor does it discuss potential limitations of typical models such as $\Pi$ equivalent transmission line model which may no longer be valid when the frequency changes significantly. %

In this paper, we aim to address these gaps.
We first discuss the frequency-dependent components of transmission line models and analyze the $\Pi$ equivalent model in detail to discuss how the definitions of ``long'' and ``short'' transmission lines change as we vary the frequency. This allows us to verify that typical modeling assumptions remain valid as we lower the frequency and to provide an analysis that extends to frequencies above 60 Hz or to the limiting case of 0 Hz.
We then use an optimization-based framework to analyze how the maximum power transfer capacity of a transmission line varies with frequency. Finally, we investigate how changing the frequency changes the admissible combinations of active and reactive power transfer.

To summarize, the contribution of this paper is to analyze the behavior of transmission lines across a range of different frequencies. Specifically, we introduce (i) a variable frequency $\Pi$ branch model with analytical validation as well as models for capacitive and inductive bus shunts, and (ii) analytical means to analyze power transfer across a line for different frequencies, including DC.

Section \ref{sec:freq_dep} establishes assumptions for frequency-dependent parameters. Section \ref{sec:pi_model} then introduces the frequency-dependent $\Pi$ branch model with variable frequency and analyzes the range of validity for the different line model approximations. The validated model is used in Section \ref{sec:powerflow} to demonstrate the power flow capability of the branches as a function of frequency. Conclusions follow in Section \ref{sec:conclusions}.

\section{Frequency-dependent model parameters}\label{sec:freq_dep}

This section discusses the frequency dependence of parameters in models for transmission lines, cables and bus shunts. We do not consider how variations in frequency would affect generators or loads, since we assume that such components would be connected to parts of the grid with standard AC frequencies. Furthermore, we omit discussions of transformers, as we assume that LFAC would most likely be designed within a single voltage level or if not, that power electronic converters would be used for voltage conversion to avoid the need for very large transformers.

\subsection{Frequency-dependent line and cable parameters}\label{sec:freq_effects}
Transmission lines and cables are traditionally modeled as circuits with four parameters: series inductive reactance $X'$, series resistance $R'$, shunt capacitive susceptance $B'$, and shunt conductance $G'$ \cite{Zaborszky1954}\cite{kundur1994}. These parameters are distributed throughout the length of the line and are given as values per unit length, denoted with the prime ($'$) symbol.

While the line parameters are typically assumed to be constant, when the system frequency varies, as in the design or operation of LFAC systems, the line parameters start to change.
Most significantly, the inductive series reactance $X'$ and capacitive shunt susceptance $B'$ depend on the product between frequency and the inductance $L'$ or the capacitance $C'$, respectively,
\begin{align}
	X'(\omega)=\omega L',\quad\quad	B'(\omega)=\omega C'.\label{eq:susceptance}
\end{align}
The series resistance, $R'$, and the shunt conductance, $G'$, do not have significant frequency dependence and are modeled as fixed for the full range of frequencies considered here.

We note that while the linear frequency dependence of the series reactance $X'$ and shunt susceptance $B'$ is by far the most dominant dependence, there are also other frequency-dependent effects.
Both the skin effect and proximity effect are dependent on frequency and contribute to lowering both the series resistance and inductance as frequency is reduced \cite{Zaborszky1954}\cite{bergen2000}. Manufacturer data for series resistance at 60 Hz and DC for typical aluminum conductor steel reinforced (ACSR) transmission lines shows the DC resistance to be about 2-3\% less than at 60 Hz \cite{generalcable2017}. In \cite{Ngo2016PowSys}, an analytical model showed the skin effect to contribute to a 0.25\% change in the resistance between these frequencies.
These effects contribute to a reduction in the series resistance and reactance as the frequency decreases, and in some cases highly accurate models may need to incorporate them. However, these effects are insignificant in typical steady state power flow calculations and are omitted from our subsequent analysis.

\subsection{Frequency-dependent bus elements}
A shunt element, capacitive or inductive, can also be directly connected to a bus. These elements are represented as a resistance and either capacitance or inductance connected in parallel between the bus and ground. %
In the same way as for the transmission line, we assume that the shunt resistance $G_{i}^\mathrm{sh} = 1/R_{i}^\mathrm{sh}$ is independent of frequency, while the shunt susceptance is given by
\begin{equation}
    B_{\mathrm{cap},i}^\mathrm{sh} (\omega)= \omega C_{i}^\mathrm{sh}
    \quad\textrm{or}\quad
    B_{\mathrm{ind},i}^\mathrm{sh} (\omega)= \frac{-1}{\omega L_{i}^\mathrm{sh}}
\end{equation}
for a capacitive shunt $B_{\mathrm{cap},i}^\mathrm{sh}$ or an inductive shunt $B_{\mathrm{ind},i}^\mathrm{sh}$, respectively, at any bus $i$.

\section{Validity of the lumped-parameter \texorpdfstring{$\Pi$}{Pi} model}\label{sec:pi_model}

In power system modeling, it is very common replace the exact transmission line model based on the distributed parameters with the lumped-parameter $\Pi$ model. This approximation is typically assumed to be valid for medium-length transmission lines. In this section, we investigate how changes in the frequency change the definition of ``medium length''.

\subsection{\texorpdfstring{$\Pi$}{Pi} equivalent branch model}
\begin{figure}%
\centering
\resizebox{0.45\textwidth}{!}{%
\includegraphics{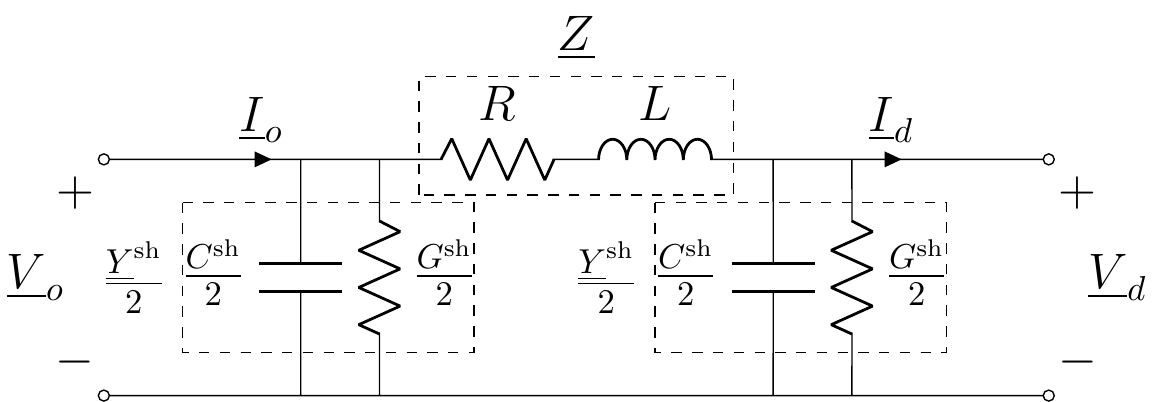}
	}
    \caption{Lumped-parameter $\Pi$ model of a branch, with series impedance $\underline{Z}$ and shunt admittance $\underline{Y}^\mathrm{sh}.$
    \vspace{-0.1in}
    }
    \label{fig:branch_pi}
\end{figure}

Transmission branches are commonly represented as a lumped-parameter $\Pi$ model, as shown in Fig. \ref{fig:branch_pi}.
From this model, the terminal quantities at the origin ($o$) and destination ($d$) sides, phasors $\underline{I}_o$, $\underline{V}_o$, $\underline{I}_d$, and $\underline{V}_d$, are given by
\begin{equation}
		\renewcommand*{\arraystretch}{1.7}
	\begin{bmatrix}
		\underline{V}_o\\
		\underline{I}_o
	\end{bmatrix}=
	\begin{bmatrix}
		1+\underline{Z}\cdot\frac{\underline{Y}^\mathrm{sh}}{2} & \underline{Z}\\
		\underline{Y}^\mathrm{sh}\left(1+\underline{Z}\cdot\frac{\underline{Y}^\mathrm{sh}}{4}\right) & 1+\underline{Z}\cdot\frac{\underline{Y}^\mathrm{sh}}{2}
	\end{bmatrix}
	\begin{bmatrix}
		\underline{V}_d\\
		\underline{I}_d
	\end{bmatrix},
	\label{eq:pi_matrix}
\end{equation}
where $\underline{Z}=R+j\omega L$ and $\underline{Y}^\mathrm{sh}=G^\mathrm{sh}+j\omega C^\mathrm{sh}$.
The lumped-parameter $\Pi$ model is obtained by approximating the lumped parameters as the distributed parameter multiplied by the length of the line, $R=\ell R',~L=\ell L',~G^\mathrm{sh}=\ell G'$ and $C^\mathrm{sh}=\ell C'$ where $\ell$ is the length of the branch.

\subsection{Exact transmission line model}
While the lumped-parameter model is a common approximation, the exact transmission line model provides the terminal quantities exactly in terms of the distributed parameters and line length \cite{bergen2000},
\begin{equation}
	\renewcommand*{\arraystretch}{1.7}
	\begin{bmatrix}
		\underline{V}_o\\
		\underline{I}_o
	\end{bmatrix}=
\begin{bmatrix}
	 \cosh(\underline{\gamma}\ell) & \underline{Z}_0 \sinh(\underline{\gamma}\ell)\\
	 \frac{1}{\underline{Z}_0}\sinh(\underline{\gamma}\ell) & \cosh(\underline{\gamma}\ell)
\end{bmatrix}
\begin{bmatrix}
\underline{V}_d\\
\underline{I}_d
\end{bmatrix}. \label{eq:abcd_l}
\end{equation}
Here, ${\underline{\gamma}=\sqrt{(R'+jX')\cdot(G'+jB')}}$ is the propagation constant, and $\underline{Z}_0=\sqrt{(R'+jX')/(G'+jB')}$ is the characteristic impedance of the line.

\subsection{Relationship between the \texorpdfstring{$\Pi$}{Pi} model and the exact model}
The lumped-parameter $\Pi$ model is an exact representation for the terminal quantities when the elements of \eqref{eq:pi_matrix} are equal to those in \eqref{eq:abcd_l}.
Setting the lumped parameters to match \eqref{eq:abcd_l} we find $\frac{\underline{Y}^\mathrm{sh}}{2}$ and $\underline{Z}$ in terms of the distributed parameters:
\begin{align}
	\frac{\underline{Y}^\mathrm{sh}}{2} &= \frac{1}{\underline{Z}_0}\tanh\left(\frac{\underline{\gamma}\ell}{2}\right), \quad
	\underline{Z} =\underline{Z}_0\sinh(\underline{\gamma\ell}).\label{eq:pi_yz}
\end{align}
If $|\underline{\gamma}\ell|\ll1$, the arguments of the $\tanh$ and $\sinh$ functions are small and the functions can be approximated as the values of their arguments. With this approximation, we have that
\begin{align}
	\frac{\underline{Y}^\mathrm{sh}}{2}&\approx\frac{1}{\underline{Z}_0}\cdot\frac{\underline{\gamma}\ell}{2}=\frac{G'\ell}{2}+j\omega \frac{C'\ell}{2}\label{eq:pi_yc_appr}\\
	\underline{Z}&\approx \underline{Z}_0\cdot\underline{\gamma}\ell=R'\ell+j\omega L'\ell.\label{eq:pi_zs_appr}
\end{align}
Thus, for sufficiently small $|\underline{\gamma}\ell|$, the lumped-parameter $\Pi$ model is a good approximation of the exact model.

\subsection{Validation of \texorpdfstring{$|\underline{\gamma}\ell|\ll1$}{|gl|} for varying frequencies}
The accuracy of the lumped-parameter $\Pi$ model depends on the quantity $\vert\underline{\gamma}\ell\vert$ having a sufficiently small magnitude. In typical power system models, the line length is the only variable quantity and the lumped-parameter $\Pi$ model is assumed to be valid for lines that are shorter than $<250$ km. However, when the frequency is no longer fixed, we need to assess the validity of the approximation as a function of both length and frequency. We expand the quantity $\vert\underline{\gamma}\ell\vert$:
\begin{align}
	\vert\underline{\gamma}\ell\vert &= \left\vert\ell\sqrt{(R'+j\omega L')(G'+j\omega C')}\right\vert \label{eq:mag_vs_omega_l}\\
	&= \ell\sqrt{L'C'}\omega\sqrt{\left(\left(\frac{R'}{\omega L'}\right)^2+1\right)^\frac{1}{2}\left(\left(\frac{G'}{\omega C'}\right)^2+1\right)^\frac{1}{2}}.\nonumber
\end{align}
At very low frequencies, the $\omega$-dependent terms of \eqref{eq:mag_vs_omega_l} approach the limit $\sqrt{(R'G')/(L'C')}$. This holds when the frequency is much smaller than the ratios $\frac{R'}{L'}$ and $\frac{G'}{C'}$, i.e.,
\begin{align}
	\vert\underline{\gamma}\ell\vert &\approx \ell\sqrt{R'G'}, \qquad\text{if } \omega \ll \frac{R'}{L'} \quad\text{and } \omega \ll \frac{G'}{C'}.\label{eq:mag_approx_lf}
\end{align}
As $\omega$ increases, the expression under the square root decreases monotonically to the limit of 1. Thus, at frequencies much larger than $\frac{R'}{L'}$ and $\frac{G'}{C'}$, the $\omega$-dependence is nearly linear and $\vert\underline{\gamma}\ell\vert$ is dominated by $\ell\omega\sqrt{L'C'}$:
\begin{align}
	\vert\underline{\gamma}\ell\vert &\approx \ell\omega\sqrt{L'C'}, \qquad\text{if } \omega\gg\frac{R'}{L'}\quad\text{and } \omega\gg\frac{G'}{C'}.\label{eq:mag_approx}
\end{align}
At intermediate frequencies, the $\omega$-dependence remains approximately linear as shown in our numerical example below, and the effect of the frequency on the $\Pi$ model approximation is comparable to that of line length.

From the analysis above, and in particular from \eqref{eq:mag_approx}, we observe that for any branch with a given length and any impedance and admittance parameters, the lumped-parameter $\Pi$ model represents the line with equal or better accuracy as the frequency decreases.

\subsection{Numerical example}\label{sec:numerical_ex}
To illustrate the effect of frequency dependence, we consider a typical three phase overhead transmission line with ACSR conductors. Manufacturer data \cite{generalcable2017} gives the following distributed parameter values:
\begin{align*}
	R'&=0.05709\quad\Omega/\mathrm{km} \quad\quad &&C'=9.497\quad\mathrm{nF}/\mathrm{km}\\
	L'&=1.214\quad\mathrm{mH}/\mathrm{km}\quad\quad &&	G'=0\quad\Omega/\mathrm{km}
\end{align*}
We assume that the origin side voltage is fixed at 1.0 p.u. and that the line is operated under surge impedance loading conditions, $\underline{I}_d=\underline{V}_d/\underline{Z}_0$, and consider frequencies from 0 to 80 Hz and line lengths from 100 to 700 km.
To demonstrate how the approximation quality varies for different frequencies, we compute (i) the value of $|\underline{\gamma}\ell|$ and (ii) the error of the lumped-parameter $\Pi$ model approximation as the difference between the terminal voltage $V_d$ obtained by solving \eqref{eq:pi_matrix} and \eqref{eq:abcd_l}, respectively. We also compare how the obtained values compare with the values for a 250 km line at 60 Hz, which has $|\underline{\gamma}\ell|=0.321$ and an error of $2.21\cdot10^{-3}$ p.u.

\begin{figure}%
	\begin{subfigure}[t]{0.5\textwidth}
		\centering
		\resizebox{0.7\textwidth}{!}{
		\small
		\def\svgwidth{0.9\textwidth}
		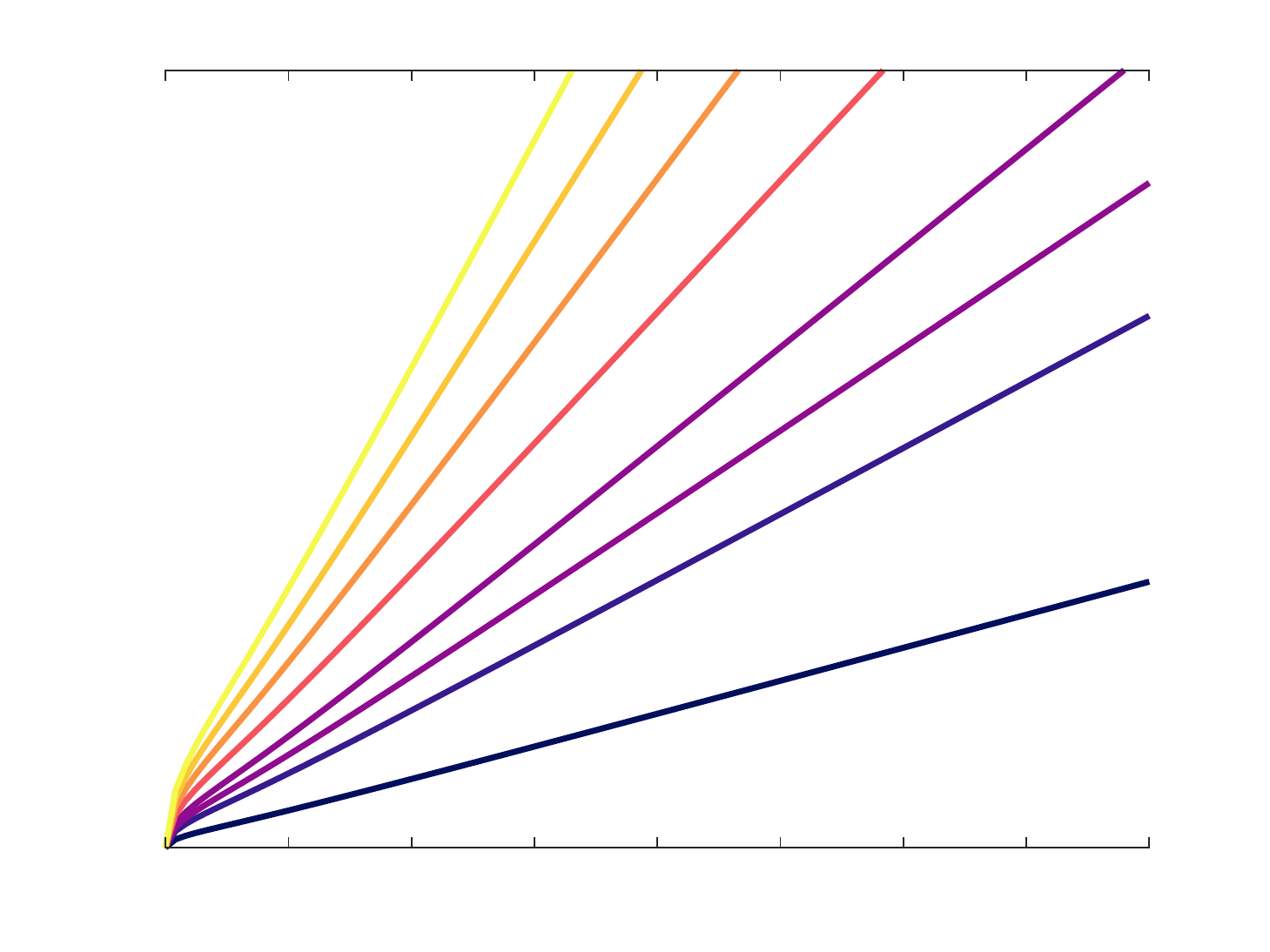
		}
		\caption{Magnitude of the characteristic impedance times length for the given parameters vs. frequency at various lengths. The validity of the $\Pi$ model approximation depends on this value being small.}
		\label{fig:pi_error}
	\end{subfigure}
	\begin{subfigure}[t]{0.5\textwidth}
		\centering
		\resizebox{0.7\textwidth}{!}{
		\small
		\def\svgwidth{0.9\textwidth}
		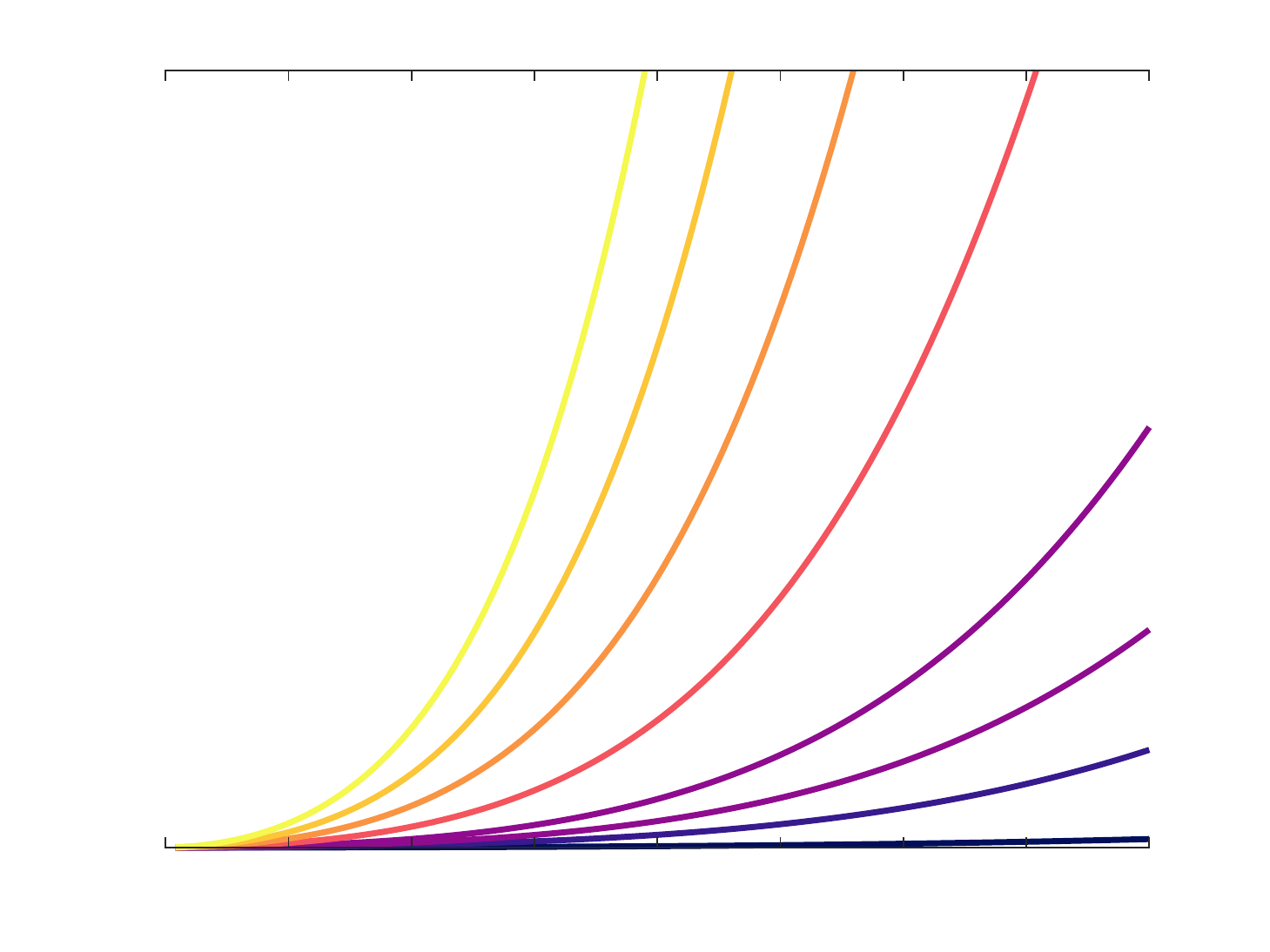
		}
		\caption{Magnitude of the error in the destination end voltage calculation using the $\Pi$ approximation vs. frequency for a line at various lengths under surge impedance loading conditions.}
		\label{fig:pi_error_sil}
	\end{subfigure}
	\caption{Frequency dependence of approximation validity and error for example line parameters at various line lengths.}
\end{figure}

The value of $|\underline{\gamma}\ell|$ is shown in Fig. \ref{fig:pi_error}, and the lumped-parameter $\Pi$ model error in p.u. is plotted in Fig. \ref{fig:pi_error_sil}. %
We observe that the magnitude of $|\underline{\gamma}\ell|$ increases linearly with frequency, which is similar to the linear dependency on the line length. As a result, the error will also depend similarly on the frequency and length, although it increases exponentially as either parameter increases.
We conclude that the approximation holds quite well for low frequencies, short lines, and combinations of the two. Specifically, comparing with the errors deemed acceptable for a 250 km line at 60 Hz, it is possible to use the lumped-parameter $\Pi$ model for lines as long as 700 km at a frequency of 18 Hz.

\section{Power flow as a function of frequency}\label{sec:powerflow}
In this section, we analyze the power flow on a transmission line as a function of frequency. %

\subsection{Power flows with frequency as a variable}
We first provide the frequency-dependent power flow equations for an AC line and discuss how the equations generalize to 0 Hz DC transmission.
\subsubsection{AC power flow with frequency as a variable}
To express the power flow on a transmission line, we write the active and reactive power injected at the origin bus, $P_o$ and $Q_o$, in terms of bus voltage magnitude and angle:
\begin{align}
  P_o&\!=\!V_o^2(G\!+\!G^\mathrm{sh})\!-\!V_oV_d\left(G\cos(\theta_{od})\!+\!B\sin(\theta_{od})\right)\label{eq:pflow}\\
  Q_o&\!=\!-V_o^2(B\!+\!B^\mathrm{sh})\!-\!V_oV_d\left(G\sin(\theta_{od})\!-\!B\cos(\theta_{od})\right).\label{eq:qflow}
\end{align}
Here, $V_o$ and $V_d$ are the origin and destination bus voltage magnitudes, and $\theta_{od}=\theta_o-\theta_d$ is the angle difference between the buses. The series conductance $G$ and the series susceptance $B$ are derived based on the lumped-parameter $\Pi$ model in the last section. Their frequency-dependent values are given by
\begin{align*}
  G(\omega)&=\frac{R}{R^2+\omega^2L^2} \quad\quad
  B(\omega)=\frac{-\omega L}{R^2+\omega^2L^2},
\end{align*}
while the capacitive shunt susceptance $B^\mathrm{sh}$ is given by $B^\mathrm{sh}(\omega)=\omega C^\mathrm{sh}$.

\subsubsection{Generalization of power flow equations to 0 Hz}
The behavior with variation of frequency continues to the limit of 0 Hz. Approaching 0 Hz, the voltage and current waveforms are sinusoidal with an increasingly long period, but in the case of zero frequency, the circuit becomes quite different. The voltage and current, defined in AC by their root-mean-square (RMS) magnitude, become constant DC waveforms. The choice of DC voltage on each conductor is non-trivial, depending on the number of conductors involved and on several factors related to insulation \cite{LARRUSKAIN20111341}. We represent the effects of these factors on the power flow with a constant $k$. For example, for a 3-conductor circuit with the DC pole to neutral voltage equal to the AC RMS phase to neutral value, this constant is $k=\frac{2}{3\sqrt{3}}$. In addition, the voltage angle difference, $\theta_{od}$, loses its significance and becomes zero. The value of $G$ becomes $\frac{1}{R}$ and both $B$ and $B^\mathrm{sh}$ become zero. This gives the power flow equations in the 0 Hz limit:
\begin{align}
  P_o^\mathrm{dc}&=k\left(V_o^2\left(\frac{1}{R}+G^\mathrm{sh}\right)-\frac{V_oV_d}{R}\right)\label{eq:pflowdc}\\
  Q_o^\mathrm{dc}&=0.\label{eq:qflowdc}
\end{align}

\subsubsection{Constraints on power flow}
The solution to the power flow equations is subject to several system constraints. First, the branch conductor has thermal limits imposed as a maximum apparent power magnitude $S^\mathrm{max}$ injected to the branch by each connected bus $i$,
\begin{equation}
	|S_i|=\sqrt{P_i^2+Q_i^2}\leq S^\mathrm{max}, \quad\forall i\in\{o,d\}.\label{eq:slim}
\end{equation}
Next, the voltage magnitude at each connected bus $i$ must also lie within the limits $V^\mathrm{min}$ and $V^\mathrm{max}$,
\begin{equation}
	V^\mathrm{min}\leq V_{i}\leq V^\mathrm{max},\quad\forall i\in\{o,d\}.\label{eq:vlim}
\end{equation}
Finally, the voltage angle difference $\theta_{od}$ across the branch has limits for system stability with a maximum angle $\theta^\mathrm{max}$,
\begin{equation}
	-\theta^\mathrm{max}\leq\theta_{od}\leq\theta^\mathrm{max}.\label{eq:anglim}
\end{equation}

\subsection{Maximum Power Transfer}
We next provide a method to assess the maximum power transfer capacity for a given transmission line as a function of frequency. %

\subsubsection{Maximizing the power flow}
To more thoroughly examine the utilization of the branch, we consider the maximum power flow that is possible, dependent on the frequency. For simplicity of presentation, we fix the origin bus voltage to 1.0 p.u. We maximize the utilization of the branch for active power flow, subject to the power flow equations and limits:
\begin{align*}
	P_o^{\max} = &\max_{\omega,V_d,\theta_{od}}\quad P_o\\
	\mathrm{s.t.}\quad &(\ref{eq:pflow},\ref{eq:qflow})&&\text{power flow equations}\\
	&\eqref{eq:slim}, \eqref{eq:vlim}, \eqref{eq:anglim}&&\text{thermal, voltage, angle limits}\\
	&V_o = 1.0&&\text{origin end voltage}
\end{align*}
Here, the frequency $\omega$ is treated as an optimization variable, and the problem can be solved using a standard non-linear optimization solver. However, it is also possible to treat $\omega$ as a parameter and perform a parameter sweep to analyze the dependence of $P_o^{\max}$ over a range of frequencies. For the latter case, the maximum power flow can be found analytically over a range of frequencies by solving the power flow equations, while assuming that different sets of constraints are active (i.e., satisfied with equality).

\subsubsection{Illustrative example}
\begin{figure}[htbp]
		\centering
		\resizebox{0.45\textwidth}{!}{
		\small
		\def\svgwidth{0.5\textwidth}
		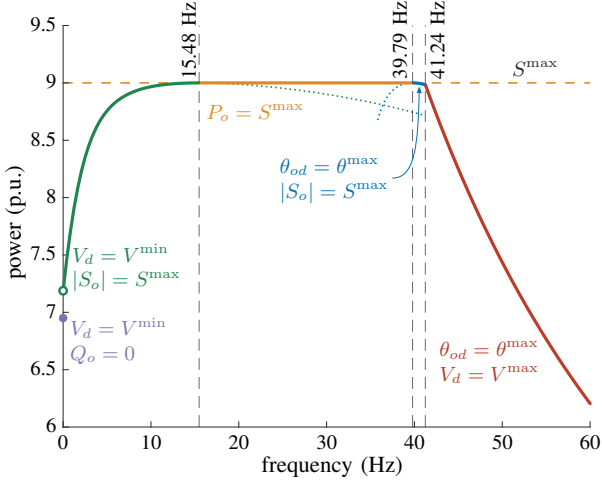
		}
		\setlength{\belowcaptionskip}{-2pt}
		\caption{Maximum active power flow on the branch, subject to the limits on angle, apparent power, and voltage magnitude.}
		\label{fig:max_Po}
\end{figure}

As an example of the solutions of this problem and their dependence on frequency, we consider the overhead line with parameters given in Section \ref{sec:numerical_ex} with $S^\mathrm{max} = 9$ p.u. apparent power limit (345 kV, 100 MVA base). The destination voltage, $V_d$ is allowed to vary between $V^\mathrm{min}=0.9$  p.u. and $V^\mathrm{max}=1.1$ p.u, and the angle limit is set to $\theta^{\max}=40^\circ$.
We solve the problem for frequencies ranging between 0 and 60 Hz and plot the results in Fig. \ref{fig:max_Po}.

We observe that there are five distinct regions:
\begin{enumerate}
    \item \emph{Angle and voltage constrained:} As the frequency decreases from 60 Hz down to 41.24 Hz, the angle limit is always active, and the destination voltage, $V_d$, is at its upper limit. Reducing the frequency has a dramatic effect in this range as the reduction in impedance allows us to transmit more power at the same angle difference.
    \item \emph{Angle and thermally constrained:} At 41.24, the apparent power $S^\mathrm{max}$ reaches its limit. The angle constraint and the destination voltage are also still at their upper bounds, $\theta_{od}=\theta^\mathrm{max}$ and $V_d = V^\mathrm{max}$. Below this frequency, $V_d$ is no longer at its limit, and the apparent power and angle are the active constraints until 39.79 Hz, at which the angle constraint becomes inactive.
    \item \emph{Thermally constrained:} In the range of 15.48 Hz to 39.79 Hz, the maximum power flow is achieved as the reactive power flow becomes zero and $P_o$ is constrained only by the apparent power limit, with $P_o=S^\mathrm{max}$.
    \item \emph{Lower voltage limit:} At frequencies below 15.48 Hz, the lower voltage limit becomes active and nonzero reactive power flows on the line. This limits the active power flow further as frequency decreases.
    \item \emph{DC power flow:} At the 0 Hz limit, the power flow is constrained by the voltage magnitude and power flow for the DC case, (\ref{eq:pflowdc},\ref{eq:qflowdc}), and we take $k=1$. This results in a slight decrease in the active power capability. We see from this example that decreasing the frequency to 53 Hz, we can achieve higher power flow than at DC.
\end{enumerate}
An important conclusion from this example is that lower frequencies may not always be optimal for achieving maximum power transfer capability. As a result, it is important to have a proper process for choosing the optimal frequency.

\subsection{Flexibility of power flow}
While changing the frequency can allow for a higher maximum power transfer, decreasing the frequency also allows for a larger set of feasible combinations of active and reactive power transfer, giving additional flexibility to the system.

\subsection{Power circles}
We can visualize the power flow flexibility by plotting the power flow equations (\ref{eq:pflow},\ref{eq:qflow}) on the axes of active and reactive power injection from bus $o$, in the form of a \emph{power circle diagram} \cite{bergen2000}. For any values of $V_d$, the solutions to the power flow equations trace out concentric circles, or ``power circles," as $\theta_{od}$ varies.
When the frequency changes, the power circle experiences both translation and scaling.
\subsubsection{Translation}
The location of the center of the power circle, ${(P_o^0, Q_o^0)}$, is given by
\begin{align}
	P_o^0&=V_o^2\left(G(\omega)+\frac{1}{R^\mathrm{sh}}\right)\label{eq:pcenterf}\\
	Q_o^0&=-V_o^2\left(B(\omega)+\omega C^\mathrm{sh}\right)\label{eq:qcenterf}
\end{align}
From \eqref{eq:pcenterf}, we observe that the power circle moves in the positive direction on the $P_o$ axis as frequency decreases. According to \eqref{eq:qcenterf}, the circle also moves on the $Q_o$ axis. As the frequency decreases to $\frac{R}{L}$, the series susceptance $B(\omega)$ decreases and circle moves upwards. Below this frequency, $B(\omega)$ increases and the circle moves downward. The shunt capacitance $C^\mathrm{sh}$ also moves the circle downward as it produces reactive power, though its contribution becomes smaller as frequency decreases. At 0 Hz, $Q_o=0$.

\subsubsection{Scaling}
The power circle also experiences scaling (i.e., a change in the radius) as the frequency varies. Its radius $r$ depends on the magnitudes of the voltages and of the admittance, i.e.,
\begin{equation}
	r(\omega)=V_oV_d\sqrt{G(\omega)^2+B(\omega)^2}\label{eq:radius}
\end{equation}
The radius grows monotonically to $\frac{V_oV_d}{R}$ as frequency decreases towards 0 Hz, where it becomes $k\frac{V_oV_d}{R}$ by (\ref{eq:pflowdc},\ref{eq:qflowdc}).

\subsection{Power circle visualization with constraints}
\begin{figure}%
		\centering
		\resizebox{0.4\textwidth}{!}{
		\small
		\def\svgwidth{0.5\textwidth}
		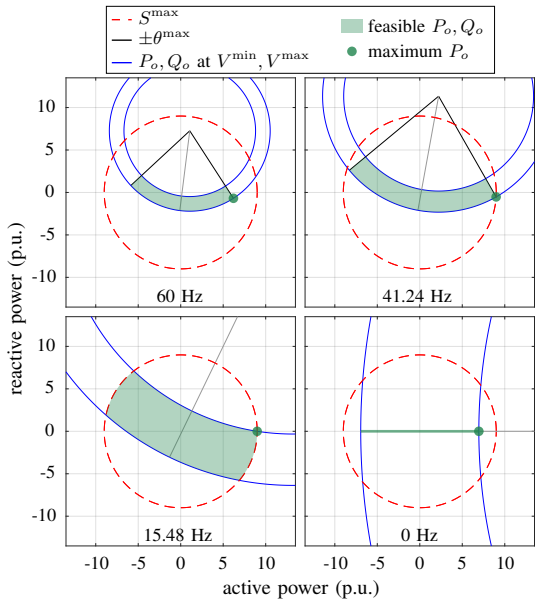
		}
		\setlength{\belowcaptionskip}{-14pt}
		\caption{Feasible active and reactive power flow on the branch from the origin bus, subject to the limits on apparent power, angle, and voltage magnitude.}
		\label{fig:circles_crossover}
\end{figure}

We next visualize the feasible active and reactive power power transfer by plotting both the power circles defined by (\ref{eq:pcenterf}-\ref{eq:radius}), as well as the limits imposed by the constraints (\ref{eq:slim}-\ref{eq:vlim}) as varying frequencies.
In Fig. \ref{fig:circles_crossover}, each subfigure shows the power circle at a different frequency, starting from 60 Hz (top left), 41.24 Hz (top right), 15.48 Hz (bottom left) and 0 Hz (bottom right). We show the power circles for $V_d=0.9$ p.u and $V_d=1.1$ p.u (blue lines) for the example line, and also include the angle constraints (black lines) and thermal limits (red lines).
By allowing $V_d$ to vary within the feasible range between 0.9 and 1.1 p.u. and considering the angle and thermal limits,
we can define a region of feasible power transmission. This feasible region is marked in green.

At 60 Hz (top left), the boundaries of the feasible region (in green) is restricted by angle limits (in black). As the frequency is lowered, the feasible region grows, and when the frequency reaches 41.24 Hz, the line can be used to its thermal potential. Throughout this transition, the maximum power transfer happens at the maximum voltage $V_d=1.1$ p.u. (i.e., the outer power circle).

As the frequency continues to decrease, the feasible region rotates clockwise, as is seen clearly at 15.48 Hz. This results in increasingly negative reactive power transfer accompanying the maximum active power transfer. The maximum power transfer happens at the lower voltage limit. Finally at 0 Hz, the feasible region is a line on the $P_o$ axis, with active power controlled only by the voltage magnitude.

Across the range of frequencies, we observe significant differences in both the power flow capacity and the flexibility. In the design of a system where frequency can be selected, it is important to understand where these frequency ranges lie. This section demonstrates that the frequency regions can be determined analytically as intersections of the power flow equations (\ref{eq:pflow},\ref{eq:qflow}) and engineering constraints (\ref{eq:slim}-\ref{eq:vlim}), dependent on branch parameters and constraint values. With these analytical solutions, appropriate selection of frequency can be combined with the utilization of existing branch conductors and design of new components to achieve optimal utilization of the transmission system.

\section{Conclusions}\label{sec:conclusions}
This paper discusses the frequency-dependent behavior of power lines and the validity of the model we use to describe them.
We first discuss how important modeling parameters depend on frequency. We then show that the lumped-parameter $\Pi$ model, which is a frequently used, approximate model of transmission lines becomes more accurate as we decrease the frequency.
Finally, we show that the maximum power transfer for an individual transmission line is typically achieved at intermediate frequencies, and that the set of feasible combinations of active and reactive power transfer also vary significantly with frequency.

The presented work provides important insights into frequency-dependent power transmission, and it demonstrates that finding the optimal frequency for power transmission is a non-trivial task. Furthermore, it lays the foundation for further developments such as variable frequency optimal power flow, which we plan to explore in future work.

\section*{Acknowledgment}
The authors would like to thank our project partners at New York Power Authority (NYPA) and in particular Greg Pedrick and Shayan Behzadirafi, as well as Giri Venkataramanan at University of Wisconsin--Madison for the discussions that helped develop and improve this work.

\bibliography{refs}{}
\bibliographystyle{IEEEtran}

\end{document}